\documentclass[preprint,showpacs,prb,floatfix,superscriptaddress,
citeautoscript,cite]{revtex4}
\usepackage{graphicx}
\usepackage{color}

\begin{document}

\title{Ag and Au Atoms Intercalated in Bilayer Heterostructures of Transition
Metal Dichalcogenides and Graphene}

\author{F. Iyikanat}
\email{fadiliyikanat@iyte.edu.tr}
\affiliation{Department of Physics, Izmir Institute of Technology,
35430, Izmir,
Turkey}

\author{H. Sahin}
\email{hasan.sahin@uantwerpen.be}
\affiliation{Department of Physics, University of Antwerp, 2610,
Antwerp, Belgium}

\author{R. T. Senger}
\affiliation{Department of Physics, Izmir Institute of Technology,
35430, Izmir, Turkey}

\author{F. M. Peeters}
\affiliation{Department of Physics, University of Antwerp, 2610,
Antwerp, Belgium}

\date{\today}


\begin{abstract}

The diffusive motion of metal nanoparticles Au and Ag on monolayer and between
bilayer heterostructures of transition metal dichalcogenides and graphene are
investigated in the framework of density functional theory. We found that the
minimum energy barriers for diffusion and the possibility of cluster formation
depend strongly on both the type of nanoparticle and the type of monolayers and 
bilayers. Moreover, the tendency to form clusters of Ag and Au can 
be tuned by creating various bilayers. Tunability of the diffusion characteristics 
of adatoms in van der Waals heterostructures holds promise for controllable growth 
of nanostructures.

\end{abstract}

\maketitle

\section{Introduction}

In recent years graphene\cite{novoselov2004electric} has become
one of the most attractive materials due to its unique properties such
as high-mobility electron transport
\cite{novoselov2005two,bolotin2008temperature}, the presence of 
room-temperature quantum Hall effect\cite{novoselov2007room},
the strong lattice structure\cite{lee2008measurement} and the extremely high
in-plane thermal conductivity.\cite{seol2010two} However, its highly active
surface and the lack of a band gap in the electronic structure are emerging
drawbacks for graphene. Recently, interests have now also
focused on other two-dimensional systems having honeycomb
structures, such as
graphane\cite{sofo2007graphane,flores2009graphene},
halogenated graphenes\cite{nair2010fluorographene, hasan-cf,
ortwin-cf, hasan-ccl},
silicene\cite{kara2012review}, III-V binary
compounds\cite{PhysRevB.52.8881}, and transition metal dichalcogenides
(TMDs)\cite{PhysRevB.8.3719, chhowalla2013chemistry, yandong1, hasan-nature}.
Recent
studies have revealed that among various monolayer structures
especially TMDs are quite promising materials for electronics and
optoelectronics applications.

Bulk TMDs have a number of exceptional properties such
as superconductivity of TaS$_{2}$ and  NbSe$_{2}$, Mott transition in
1T-TaS$_{2}$ and the presence of charge density wave in TiSe$_{2}$.
\cite{sipos2008mott, jishi2008electronic} It was also reported that
the electrical and optical properties of TMDs are dramatically altered with
the number of layers\cite{splendiani2010emerging,PhysRevB.88.075409,
hasan-wse2}. Although bulk hexagonal TMDs possess an indirect band gap,
mono-layer TMDs exhibit a direct band gap which is crucial for optoelectronic
devices, sensors and catalysts. In addition, n-type and
p-type field-effect-transistors (FETs) based on monolayer and
multilayer TMDs have been
investigated.\cite{das2012high,radisavljevic2011single,
fang2012high} It was also reported that many monolayer 2D crystals are 
reactive and segregation may occur
easily.\cite{geim2011random} Therefore, the investigation of bilayer
structures which are chemically more stable than monolayer structures
is of vital significance. Recent studies have shown that synthesis of
heterostructures made of combinations of different TMD single layers,
graphene, fluorographene and hexagonal-BN (hBN) is experimentally
achievable.\cite{geim2013van,terrones2013novel, haigh2012cross,
britnell2012field, dean2012graphene,georgiou2012vertical} Since TMDs and  
other two-dimensional structures have a lot of diverse monolayer structures, 
when they are combined together they are expected to exhibit very different 
properties. \cite{yandong,xinru}

Since the intercalation and migration of foreign atoms is inevitable
during the formation of such lamellar materials and heterostructures,
the investigation of the diffusion characteristics of various impurities is
essential. Early studies revealed that alkali-metal doping of bulk
TiSe$_{2}$ and MoS$_{2}$ can be utilized as an efficient way to tune
the Fermi level.\cite{joensen1986single,jaegermann1994photoelectron}
The electrical conductivity of MoS$_{2}$ can be altered by
substitutional doping.\cite{tiong2001electrical,PhysRevB.78.134104}
Furthermore, decoration of the surfaces of few layer TMDs by metal
nanoparticles like Au, Ag and Pt may provide p- and
n-type doping.\cite{rao2012recent,shi2013selective,kim2013enhanced,duygu}
The metal-atom adsorbed TMDs find their application in various areas
including energy storage\cite{matte2012synthesis},
photonics\cite{huang2010actively,qian2012atomically},
biosensing\cite{he2012graphene}, and catalysis.\cite{lu2010platinum}
In a recent study, it has been demonstrated that, the 
presence of various impurities at the interface between MoS$_{2}$/graphene/hBN 
and WS$_{2}$/graphene/hBN heterostructures may modify the mobility of charge 
carriers.\cite{Kretinin} It was also reported that 
contamination and migration of various molecules are inevitable during the 
formation of graphene based heterostructures and trapped hydrocarbons 
segregate into isolated pockets, leaving the rest of the interface atomically
clean. \cite{Haigh} In addition, it was found that, attached metal nanoparticles
on 
TMDs/Graphene stacks can be suitable for enhanced optoelectronic
properties.\cite{britnell2013strong, Sachs}

Despite some recent studies on adatom adsorption on various
TMDs, intercalation and migration of foreign atoms in heterostructures
have not been investigated. In this study, using density functional
theory based electronic structure method, we investigate the diffusion
characteristics of heavy metal atoms (Au and Ag) on monolayers and
intercalated in such bilayer heterostructures. The paper is organized as follows: 
In Sec. \ref{sec2} we give details of our computational methodology.
In Sec. \ref{sec3} the energetics of the metal atoms Ag and Au on monolayers of
graphene
and TMDs are presented. In Sec. \ref{sec4} diffusion characteristics
of those metal atoms inside bilayer heterostructures are shown and in Sec. \ref{sec5} 
we summarize and conclude our results.

\section{Computational Methodology}\label{sec2}

To determine ground state atomic structures and migration
characteristics
of monolayers and their bilayer heterostructures, 
first-principles calculations were performed using density functional
theory (DFT) with a plane-wave basis set as implemented in the Vienna
ab initio simulation package (VASP).\cite{PhysRevB.54.11169} 
For the exchange correlation function generalized
gradient approximation (GGA) of Perdew, Burke and
Ernzerhof\cite{PhysRevLett.77.3865} was used together with
the van der Waals
correction.\cite{grimme2006semiempirical} Spin-unpolarized
calculations were carried out using projector-augmented-wave
potentials (PAW). The plane-wave basis set with kinetic energy cutoff
of 500 eV was used.

A 4$\times$4 and 3$\times$3 hexagonal supercells of single layer and
bilayer structures are employed to model diffusion paths of metal
atoms, respectively. The k-point samples were 3$\times$3$\times$1 for
these supercells. It is calculated that a 12 \AA\ of 
vacuum space is enough to hinder long-range dispersion forces between 
two adjacent images in the supercell. Lattice constants and total
energies were computed with the conjugate gradient method, where
atomic forces and total energies were minimized. The convergence
criterion of our calculations for ionic relaxations is $10^{-5}$ eV
between two consecutive steps. The maximum Hellmann-Feynman forces
acting on each atom were reduced to a value of less than $10^{-4}$
eV/\AA. All components of the external pressure in the unit cell
was held below 1 kbar. For the electronic density of states Gaussian
smearing was used with a broadening of 0.1 eV. The non-local
correlation energies were determined by employing density functional
theory plus the long-range dispersion correction (DFT+D2)
method.\cite{grimme2006semiempirical} The values of 24.670, 5.570,
12.640, 1.750, 24.670, 40.620, and 40.620 are used as the C$_{6}$
coefficient of Mo, S, Se, C, Ag, Au, and W atoms, respectively. On the
other hand, values of 1.639, 1.683, 1.771, 1.452, 1.639, 1.772, and
1.772 are used for the vdW radius of Mo, S, Se, C, Ag, Au, and W atoms,
respectively. These C$_{6}$ coefficients and vdW radii were 
determined from previous studies. \cite{grimme2006semiempirical, amft}
It is essential that long-range dispersion correction to 
the interlayer force is taken into account in order to obtain reliable layer-layer 
distances and electronic properties of the heterobilayers.

To study the adsorption and diffusion of metal atoms in these systems, total
energies of monolayer TMDs (graphene) with adatom were calculated for 19 (12) different points which
include the high symmetry points (H, M, B, and T). Total energies of bilayers 
with adatom were calculated for only 4 high symmetry points. In these different
points, 
the adatom-surface distance was fully relaxed while the position of the adatoms parallel to 
the surface was kept fixed. To obtain accurate diffusion characteristics 
of heavy atoms on monolayer TMDs and graphene, calculations were performed using 4x4 supercells. 
In these calculations, first and second nearest neighbors of the adatom were fully relaxed 
while all the rest was kept fixed. On the other hand, in the bilayer calculations, all 
atoms of bilayers were free to move in all directions. Binding energies were
calculated for the most favorable adsorption sites. These binding energies were
calculated from the 
expression $E_{B} = E_{Monolayer} + E_{A} - E_{Monolayer+A}$, where E$_{B}$ is the binding 
energy of the metal atoms on the TMDs or graphene, E$_{Monolayer}$ is the energy of
monolayer TMDs or graphene, E$_{A}$ is the energy of metal atoms,
E$_{Monolayer+A}$ is the total energy of the metal atom-monolayer system. 

We have chosen the following convention to define the diffusion energy 
barrier. The binding energy of an adatom was calculated at all high-symmetry points 
of the TMDs and the graphene surface. Since adatoms follow the lowest energy path, the difference 
of the energies between the most favorable site and the second most favorable site was 
considered as the diffusion energy barrier.

In order to obtain the correct value of the charge transferred between the metal
atoms, the TMDs and the graphene, Bader charge analysis was performed. Rather
than electronic orbitals in Bader methodology, charge partitioning is
based on electronic charge density and therefore it is highly
efficient, it scales linearly with the number of grid points, and is
more robust than other partitioning schemes.\cite{henkelman2006fast,
tang2009grid, sanville2007improved}

\section{Diffusion of Heavy Atoms on Monolayers of TMDs
and graphene}\label{sec3}

So far, experiments have revealed that similar to graphene many of the 
TMDs (MoS$_{2}$, MoSe$_{2}$, WS$_{2}$, WSe$_{2}$, ReS$_{2}$) have a
lamellar crystal structure where the layers are held together by weak
van der Waals forces, while intra-layer metal-chalcogen bonds have
strong covalent character. Due to the weakly (vdW) bonded layered
structure of graphene and TMDs either diffusion of foreign atoms on
their surface or easy intercalation into the layers may take place.
Moreover, these foreign atoms may provide new functionalities to the TMDs
and to the heterostructures made of various layered materials. Therefore,
understanding how foreign atoms adsorb, migrate and
intercalate are of highly importance.

First we calculate the binding and migration of Ag and Au atoms on the surface
of monolayer TMDs and graphene. For a hexagonal primitive unitcell of
MoS$_{2}$, MoSe$_{2}$, WS$_{2}$ and graphene the lattice parameters
are calculated to be 3.190, 3.316, 3.179 and 2.468 \AA, respectively.
For calculations of adsorption and migration of metal atoms 4$\times$4
supercells were used in order to limit Au-Au (and Ag-Ag) interaction between
adjacent cells. As shown in Fig. \ref{structures} for a metal atom
adsorbed on TMDs and graphene, there are four possible adsorption
sites: top of carbon or chalcogen atom (T), bridge site on C-C
or M-X bond (B), on top of the center of hollow site (H) and on top of metal
atom Mo or W (M).

Diffusion and mobility of impurities on a two-dimensional crystal surface can
be described by quantities such as activation energy ($E_{a}$) and jump
probability ($P$) from the binding site over the lowest-barrier path at room
temperature. The $P$ value is a measure of the possibility of propagation by
overcoming the energy barrier among the possible adsorption sites. Jump
probability from one lattice site to another one can be calculated by using the
formula $P \approx e^{-E_{a}/k_{B}T}$ where $k_{B}$ is the Boltzmann constant and
it increases exponentially with increasing temperature. Here $E_{a}$ is the
activation energy which is equal to the difference of the energy of the two
lowest energy states.


Our calculated adsorption sites, binding energies, vertical adsorption position,
charge transfer, energy barrier and jump probability ($P$) values are listed in
Table \ref{table1}. We see that the bonding of an Ag atom to graphene occurs at
the B site and the H site is the least favorable adsorption site. In accordance
with the previous DFT study, we found that the binding energy of Au on graphene
is almost twice the binding energy of Ag on graphene. The Ag atoms lose (while
the Au atoms gain) a small amount of charge when they bind to the graphene
surface.\cite{Nakada} On the surface of TMDs, the most favorable bonding site
for Ag atoms is the H site and the next largest adsorption energy is found for
the M site. It is reasonable to assume that metal atoms diffuse through these
two favorable adsorption sites. Therefore the energy difference between these
two lowest-energy sites can be regarded as
the diffusion barrier. Since the energy barrier for Ag between H and
M sites is high ($\sim$50 meV) diffusion through these symmetry
points may not occur at low temperatures. It appears from Fig. \ref{ag-hesap} 
that Ag atoms on graphene migrate through the B and T sites with almost 
zero ($\sim$2 meV) energy barrier, while migration on TMDs (MoS$_{2}$, 
MoSe$_{2}$ and WS$_{2}$) occurs through H and M sites. As shown in Table
\ref{table1}, room temperature $P$ values of Ag atoms are 0.129, 0.182, 0.072 and
0.926 for MoS$_{2}$, MoSe$_{2}$, WS$_{2}$ and graphene, respectively. Thus
diffusion of the Ag atoms on graphene occurs much more easily than on TMDs.


Binding energy of a single Au atom on graphene and TMDs is found to be $\sim$300 meV
larger than that of Ag. Interestingly, on MoSe$_{2}$, Au atoms prefer bonding
on the 
H site like Ag atoms, while the top of a sulfur atom (T site) is the most
preferable
site on MoS$_{2}$ and WS$_{2}$ layers. The distinctive behavior of the Au atoms can
be explained by the distinctive characteristic of the S-Au bond. Although the
chemical properties of chalcogen atoms S and Se are quite similar, as shown by
Yee et al.\cite{Yee}, the strength of the S-Au bond is slightly weaker than
the Se-Au
bond. Our calculations reveal that the binding energy of an Au
atom on MoSe$_{2}$ surface is higher than the binding energy of an Au atom on
MoS$_{2}$ and WS$_{2}$ surfaces. Interestingly, strong Au-S bonds require 
bonding on T site while the Au-Se bond results in H-site adsorption. As shown in
Fig.\ref{au-hesap}, as in the Ag adsorption case, MoS$_{2}$ and WS$_{2}$
surfaces have almost identical characteristics for Au adsorption at all high
symmetry points. In addition, the migration barrier seen by Au atoms is
$\sim$30 meV larger than the energy barrier of Ag atoms on graphene.
Therefore, compared to Ag on graphene, the smaller jump probability of Au atoms
on graphene implies relatively slow and smaller nucleation of Au clusters on
graphene which agrees with experimental results.\cite{Subrahmanyam}

\begin{table*}
\footnotesize
\caption{\label{table1} Several adsorption characteristics of adatoms on 
monolayer substrates. Binding site and $\Delta\rho$ denote the energetically 
most favorable position and change in the charge of adatom, respectively. 
Binding energy, height (relative to surface), energy barrier (at high symmetry 
sites, relative to the binding site) and jump probability ($P$) of the most
favorable 
sites of Au and Ag atoms on the TMDs and graphene surface are listed.}
\begin{tabular}{lcccccccc}
\hline\hline
         & Ag/MoS$_2$ & Ag/MoSe$_2$ & Ag/WS$_2$ & Ag/Graphene   \\
\hline
Binding Site     & H & H & H & B \\
Binding energy (eV)  & 0.943 & 0.928 & 0.845 & 0.360 \\
Height (\AA)  & 1.97 & 2.07 & 2.04 & 2.57 \\
$\Delta\rho = \rho_{f} - \rho_{i}$ (e) & -0.30 & -0.18 & -0.23 & -0.16 \\
Energy Barrier (meV)  & 53(M), 146(B), 228(T) & 44(M), 139(B),
241(T) & 68(M), 111(B), 181(T) & 2(T), 57(H) \\
Jump probability ($P$) & 0.129 & 0.182 & 0.072 & 0.926 \\
\hline
         & Au/MoS$_2$ & Au/MoSe$_2$ & Au/WS$_2$ & Au/Graphene   \\
\hline
Binding Site & T & H & T & T \\
Binding energy (eV) & 1.161 & 1.232 & 1.146 & 0.633 \\
Height  (\AA)  & 2.26 & 2.03 & 2.27 & 2.56  \\
$\Delta\rho = \rho_{f} - \rho_{i}$ (e) & 0.01 & 0.08 & 0.06 & 0.22 \\
Energy Barrier (meV) & 136(M), 23(H), 46(B) & 103(M), 72(B), 68(T) &
180(M), 43(H), 55(B) & 65(H), 26(B) \\
Jump probability ($P$) & 0.411 & 0.072 & 0.189 & 0.366 \\
\hline\hline
\end{tabular}
\end{table*}

Calculated room temperature $P$ values of an Au atom on 
MoS$_{2}$, MoSe$_{2}$, WS$_{2}$ and graphene surfaces are 0.411, 0.072, 
0.189 and 0.366, respectively. Au atom has the maximum $P$ value on a 
MoS$_{2}$ surface, whereas it has the minimum $P$ value on a MoSe$_{2}$ 
surface. $P$ values of Au atoms on a MoS$_{2}$ surface is higher than that 
of Ag atoms. This result is consistent with the previous experimental 
study where they found that the cluster diameter of Au and Ag atoms on 
MoS$_{2}$ were found to be 14 and 6 nm, respectively. \cite{Bolla} 
Other experimental work showed that cluster diameters of Au and Ag
atoms on WS$_{2}$ are 19-20 and 5 nm, respectively.\cite{matte2012synthesis} In
agreement with this experimental study, our calculations show that the $P$ value
of the Au atoms on a WS$_{2}$ surface is higher than that of Ag atoms.

\section{Diffusion of Heavy Atoms Intercalated Between Van der Waals
Bilayers}\label{sec4}

The vertical stacking of graphene and other two-dimensional atomic
crystals allows for the combination of different electronic
properties. Recent
studies have shown that, electronic and optical properties of
TMDs can be altered dramatically by forming such
heterostructures.\cite{geim2013van,komsa2013electronic} It appears
that the intercalation and contamination of foreign atoms and
functional groups at the interface of these heterostructures are 
unavoidable. As shown by recent experimental study, 
heavy atoms such as Au and Ag are quite mobile on TMD
surfaces and they form clusters. \cite{Gong} Therefore, understanding 
the diffusion characteristics of heavy atoms at the interface of 
these heterostructures is of importance for the ongoing research on 
heterostructure devices.

In this section, we investigate the diffusion
characteristics of Au and Ag atoms intercalated in MoS$_2$/Graphene, MoS$_2$/MoS$_2$, 
MoS$_2$/MoSe$_2$ and MoS$_2$/WS$_2$ heterostructures.

\subsection{Diffusion through MoS$_2$/Graphene Bilayer Heterostructure}

First, we start with the heterostructure MoS$_{2}$/Graphene. However, the 
determination of the most favorable atomic structure of MoS$_{2}$/Graphene 
heterostructure is more complicated due to the lattice mismatch. Our
calculations showed that bare DFT calculations are not capable of finding the
ground state ordering of graphene placed on MoS$_{2}$ due to the presence of
many local minima corresponding to metastable states. It was seen that
low-temperature (100 K) molecular dynamic calculations are able to avoid local
minima allowing to obtain the relaxed geometric structure of MoS$_{2}$/Graphene 
as shown in Fig. \ref{structures}(c). Due to the lack of symmetry in the
MoS$_{2}$/Graphene structure, twenty-seven inequivalent adsorption sites are
considered for each Ag (and Au) atom adsorption. Table \ref{table-doping} lists
the intercalation energies obtained by using a 3$\times$3 unit cell of the most
favorable site for each configuration. Here the intercalation energy is defined
by the expression 

\begin{equation}
E_{int}=E_{Hetero+Ag(Au)}-E_{Hetero}-E_{Ag(Au)},
\end{equation} 
where $E_{int}$ is the intercalation energy of Ag(Au) atom, $E_{Hetero+Ag(Au)}$
is the energy of the metal-atom heterobilayer system, $E_{Hetero}$ is the energy of
heterobilayers and $E_{Ag(Au)}$ is the energy of the adatom. Due to the large binding
energies of adatoms on MoS$_{2}$, all the equilibrium geometries are ruled by 
MoS$_{2}$ except for the MoS$_{2}$/Au/MoSe$_{2}$ structure where Au's
binding is larger with the MoSe$_{2}$ layer.

Our calculations revealed that binding energies of Ag and Au atoms on
TMD substrates are larger than that on graphene and Au atoms are more
stable than Ag atoms on these monolayers. For the simulation of the MoS$_2$/Graphene
heterostructure we considered 4$\times$4 and 3$\times$3 unit cells for graphene
and MoS$_2$ layers, respectively. After full optimization, the lattice
parameter of
the heterostructure is calculated to be 9.791 \AA. For this supercell the
maximum
lattice mismatch is $\sim$3$\%$. 

The adsorption energies and the equilibrium geometries were calculated by
considering 27 different adsorption sites for each Ag and Au atoms at the
MoS$_2$/Graphene interface. It is seen that the most favorable site of Ag atoms
on MoS$_{2}$ is changed from H to M site upon the addition of the graphene
layer.
Energy barrier of Ag atoms is enhanced from 2 to 43 meV with graphene
layer. Although the energy barriers seen by Au atom on  MoS$_2$ are
significantly enhanced by the presence of the graphene layer, the T site of 
MoS$_2$
remains as the most favorable adsorption site of Au. 

Furthermore, we show 2D plots of the energy barriers seen by Ag and Au atoms
at the MoS$_2$/Graphene interface in Fig. \ref{G/MoS2-Ag}. It is clear
that Ag
and Au atoms have entirely different diffusion characteristics
at the MoS$_2$/Graphene
interface: (i) diffusion of Ag atoms are much easier while Au atoms may
diffuse at high temperatures, (ii) diffusion of an Ag atom occurs through the
bridge sites of the top-lying graphene layer, (iii) diffusion of Au atom may occur
from one top-sulfur site to another one through the top site of Mo atoms. Here,
similar characteristic diffusion behavior of heavy atoms can be expected for
bilayer heterostructures of WS$_2$/Graphene as well.

\begin{table*}
\footnotesize
\caption{\label{table-doping} Several adsorption characteristic of adatoms between 
bilayer heterostructures. Binding site and $\Delta\rho$ denotes the energetically 
most favorable position and change in the charge of adatom, respectively. LL Distance 
gives the interlayer distance. Intercalation energy, height above MoS$_{2}$, energy 
barrier (at high symmetry sites, relative to the binding site) and jump probability ($P$) 
of the most favorable sites of Au and Ag atoms between bilayer heterostructures are listed. }
\begin{tabular}{lcccccc}
\hline\hline
         & MoS$_2$/Ag/MoS$_2$ & MoS$_2$/Ag/MoSe$_2$ &
MoS$_2$/Ag/WS$_2$ &
MoS$_2$/Ag/Graphene  \\
\hline
Binding Site     & H & H & H & M \\
Intercalation Energy (eV) & -1.461 & -1.395 & -1.319 & -1.344 \\
Height above MoS$_2$ (\AA) & 1.85 & 1.71 & 1.85 & 1.61 \\
$\Delta\rho = \rho_{f} - \rho_{i}$ (e) & -0.42 & -0.41 & -0.41 & -0.47
\\
LL Distance (pristine, with Ag)  & 3.08, 3.70 & 3.10, 3.60 & 3.09,
3.71 & 3.32, 3.86 \\
Energy Barrier (meV) & 222(M), 282(B) & 206(M), 308(B) & 243(M),
295(B) & 43(H), 196(T) \\
  & 222(T) & 274(T) & 340(T) & \\
Jump probability ($P$) & 2$\times$10$^{-4}$ & 3.5$\times$10$^{-4}$ &
8$\times$10$^{-5}$ & 0.189 \\
\hline
         & MoS$_2$/Au/MoS$_2$ & MoS$_2$/Au/MoSe$_2$ &
MoS$_2$/Au/WS$_2$ &
MoS$_2$/Au/Graphene  \\
\hline
Binding Site     & B & H & H & T \\
Intercalation Energy (eV) & -1.645 & -1.562 & -1.504 & -1.541 \\
Height above MoS$_2$ (\AA)  & 2.07 & 1.79 & 1.85 & 2.07 \\
$\Delta\rho = \rho_{f} - \rho_{i}$ (e) & -0.15 & -0.13 & -0.18 & -0.21
\\
LL Distance (pristine, with Au) & 3.08, 4.12 & 3.10, 3.72 & 3.09,
3.74 & 3.32, 4.35 \\
Energy Barrier (meV)  & 12(H), 96(M), 95(T)& 118(M), 75(B),  103(T)& 78(M),
24(B), 164(T) & 468(H) 326(M) \\ 
Jump probability ($P$) & 0.628 & 0.055 & 0.395 & 3$\times$10$^{-6}$ \\
\hline\hline
\end{tabular}
\end{table*}

\subsection{Diffusion through MX$_2$/MX$_2$ Bilayer Heterostructures}

Next we investigate the diffusion and energy barriers which are seen by Ag
and Au atoms sandwiched in between single layers of TMDs. There are two possible
stacking types for bilayer TMDs; AA and AB stacking. In the AA stacking,
the chalcogen atom (S or Se) in one layer is on top of the chalcogen atom (S or
Se) of the other layer. On the other hand, in AB stacking, chalcogen atom (S
or Se) in one layer is on top of the metal atoms (Mo or W) of the other layer.
Our calculations show that for the considered bilayer 
heterostructures of TMDs, the AB stacking is the ground state geometry.

\textit{MoS$_2$/MoS$_2$ Bilayers:} The lattice parameter of AB-stacked bilayer
MoS$_2$/MoS$_2$ structure was found to be almost the same as of monolayer
MoS$_2$. It
is seen that the Ag-MoS$_2$ distance decreases due to the presence of the upper
MoS$_2$ layer.
The layer-layer distance of bilayer MoS$_{2}$ increases from 3.08 to 3.70 \AA\
upon single Ag intercalation.  On the other hand, single Au atom causes a larger
layer
separation (from 3.08 to 4.12 \AA\/) than Ag atoms. Our calculations revealed that
the most preferable site of the Ag atom remains the H-site, even after the
addition of the second layer. Our charge analysis showed that metal atoms tend
to
donate a larger amount of charge when they are intercalated in bilayers. The
amount of charge transfer becomes 0.42$e$ and 0.15$e$ for Ag and Au atoms,
respectively. However, it is worthwhile to note that the preferable adsorption
site for
Au atoms changes from T to B site. Au atoms tend to bind with two 
S atoms that are in two different layers. For this reason, the B-site is the
most favorable 
site.

The minimum diffusion barrier of Ag atom is increased 
by $\sim$ 170 meV after the addition of the second layer.
Interestingly 
diffusion barriers of the Ag atom between bilayer MoS$_2$ from H to M and H to T
sites are the same. The minimum diffusion barrier of Au atom is reduced 
by half when a second layer is added. These results
show that the diffusion behavior of Ag and Au atoms between bilayer MoS$_{2}$
are
completely different. Ag atoms are not likely to form clusters inside the
bilayer MoS$_{2}$. On the other hand, Au atoms may form clusters between two
MoS$_{2}$ layers and the clusterization weight of them is larger 
than that on single layer of MoS$_{2}$.

\textit{MoS$_2$/MoSe$_2$ Bilayers:} For the geometry optimization of the
MoS$_2$/MoSe$_2$ bilayer structure we performed full atomic and lattice parameter
relaxation and therefore internal strain due to lattice mismatch is lowered
to less than 1 kB. The lattice parameter of MoS$_2$/MoSe$_2$ is
calculated to be 3.252 \AA. For this lattice parameter, the maximum lattice
mismatch is $\sim$2$\%$ which is in the range of experimentally available
strain values. When Au and Ag atoms are intercalated between MoS$_2$/MoSe$_2$,
they
show the same trend; their height above the MoS$_{2}$ surface are fairly close
(1.7 \AA\ for Ag and 1.8 \AA\ for Au) and the layer-layer distance increases
from 3.1 to 3.6 and 3.7 \AA\ for Ag and Au, respectively. H-site is the most
favorable site for Ag and Au atoms when they are located between MoS$_2$/MoSe$_2$ 
bilayer. When Ag
atom is inserted in MoS$_2$/MoSe$_2$, it donates 0.41$e$ to neighboring
TMD layers. It appears that the majority of the charges are transfered to the
MoS$_2$ layer. Similarly the Au atom loses 0.13$e$ charge when inserted between
MoS$_2$/MoSe$_2$. The minimum diffusion barrier of Ag atom is
increased by $\sim$ 150 meV when the second layer is present. But the 
second favorable site does not change with the addition of the MoSe$_2$ layer
(M). The minimum diffusion barrier of Au atom increases from 23 to 75 meV with
MoSe$_2$ addition. Second favorable site of Au atom changes from H to B site
when MoSe$_2$ layer is added. Similar to MoS$_{2}$/MoS$_{2}$, the jump probability
of Ag atom becomes zero with the addition of MoSe$_{2}$ as a second layer. The
$P$ value of Au atom decreases from 0.411 to 0.055 with the addition of MoSe$_{2}$
layer. Therefore, the formation of Ag cluster between MoS$_2$/MoSe$_2$ bilayer
heterostructure is hindered by the presence of the second layer. The nucleation of 
Au clusters may happen with a slower nucleation rate in MoS$_2$/MoSe$_2$ as compared 
to monolayer MoS$_2$.

\textit{MoS$_2$/WS$_2$ Bilayers:} The lattice mismatch between MoS$_2$ and
WS$_2$ is negligible and the lattice parameter of the bilayer structure was
calculated to be 3.182 \AA. Both Au and Ag atoms prefer to be bonded on the H
site 1.85 \AA\ away from the surface of MoS$_2$. When Ag and Au atoms
are inserted
between MoS$_2$/WS$_2$, the interlayer distances increase from 3.09 to 3.71 and
3.74 \AA, respectively. We found that the Ag atom donates 0.41 $e$ to the
surrounding MoS$_2$/WS$_2$ bilayer structure whereas the amount of charge
transfer is 0.18 $e$ for Au. The minimum diffusion barrier of the Ag atom
increases $\sim$ 190 meV with the addition of WS$_2$ layer, but it does not
change the second favorable site of Ag atom (M). It is seen that the addition of
the WS$_2$ layer does not have any significant effect on the minimum diffusion
barrier of the Au atom.

In this case, the change of the $P$ value of the Ag atom shows the same trend with
bilayer
MoS$_2$ and MoS$_2$/MoSe$_2$. In these cases, the $P$ value of Ag atom 
becomes zero with the addition of the second layer (WS$_2$ in this case). On the
other 
hand, addition of a WS$_2$ layer does not change the $P$ value of
the Au atom
between MoS$_2$/WS$_2$, significantly. As a result, Ag atoms are not likely to form clusters 
when intercalated in MoS$_2$/WS$_2$, while the clusterization probability of Au
atoms does almost not 
change with WS$_2$ layer addition.

\section{Conclusion}\label{sec5}

We found that the binding energy of Ag and Au atoms is 
maximum when they are placed on MoS$_{2}$ and MoSe$_{2}$, respectively and the
binding energy of these atoms on graphene is lower than that on TMDs. Ag
atoms 
prefer to bind at the H site on TMDs, while the H and the T sites are
most favorable adsorption sites for Au atoms. Just as for the 
Ag atom, Au atom binds to H site on MoSe$_{2}$. However, Au atom binds to T
site on MoS$_{2}$ and WS$_{2}$. When Ag atom adsorbs on these monolayers, it
tends 
to donate charge, whereas Au atom tends to receive charge from these monolayers.
Calculated $P$ values revealed that, clusterization tendency of Ag and Au are
the largest on graphene and MoS$_{2}$ layers, respectively.

Addition of a second layer has a considerable effect on the diffusion
phenomena of Ag and Au atoms. Ag atom binds to H site of all TMDs heterostructures,
whereas it binds to M site of the MoS$_2$/Graphene heterostructure. When an Au atom is
inserted between MoS$_2$/MoSe$_2$ and MoS$_2$/WS$_2$ heterostructures, H site is
the energetically most favorable site. B and T sites are the most favorable sites for the Au
atom when it is placed between MoS$_2$/MoS$_2$ and MoS$_2$/Graphene
heterostructures. The clusterization probability of Ag atom becomes zero when
TMDs are added as a second layer. On the other hand, the $P$ value of Ag atom
increases with the addition of a graphene layer. Addition of a graphene layer
hinders diffusion of Au atom. WS$_{2}$ addition as a second layer, does not make
a significant effect on the cluster formation probability of Au atom. Addition
of MoSe$_{2}$ as a second layer reduces the clusterization of Au, whereas the addition
of a second MoS$_{2}$ layer increases it. Our calculated intercalation energies
showed that, when Au and Ag atoms are inserted between MoS$_2$/Graphene,
MoS$_2$/MoS$_2$, MoS$_2$/MoSe$_2$ or MoS$_2$/WS$_2$, they are energetically more
stable as compared to the case where Ag and Au atoms are adsorbed on the
monolayer substrates. The energy barrier for Au atoms is lower than
the one for Ag atoms, when they are inserted between TMD heterostructures.
On the other hand, the energy barrier of Ag atoms is lower than the energy
barrier of Au atoms when they are inserted between MoS$_{2}$/Graphene.
Tunability of diffusion barriers seen by adatoms in van der Waals
heterostructures makes these materials potential hosts for controlled growth of
nanoclusters.
\linebreak
  
\begin{acknowledgments} 
This work was supported by the Flemish Science Foundation (FWO-Vl)
and the Methusalem foundation of the Flemish government.
Computational resources were provided by TUBITAK ULAKBIM,
High Performance and Grid Computing Center (TR-Grid e-Infrastructure),
and HPC infrastructure of the University of Antwerp (CalcUA) a
division of the Flemish Supercomputer Center (VSC), which is funded by
the Hercules foundation. H.S. is supported by a FWO Pegasus Marie
Curie Fellowship. F.I. and R.T.S. acknowledge the support from 
TUBITAK project 111T318.
\end{acknowledgments}

\begin{newpage}

Fig1: (Color online) Possible adsorption sites on (a) TMD and (b) Graphene.
Top (left) and side (right) views of bilayer heterostructures of
(c) TMD/Graphene and (d) TMD/TMD.
 
Fig2: (Color online) Diffusion paths of Ag on Graphene, MoS$_{2}$, 
MoSe$_{2}$, and WS$_{2}$. (see Fig.\ref{structures} for sites)

Fig3: (Color online) Diffusion paths of Au on Graphene, MoS$_{2}$, 
MoSe$_{2}$, and WS$_{2}$. (see Fig.\ref{structures} for sites)

Fig4: (Color online) Contour plots of the energy barriers (in meV units) seen 
by Ag (upper panel) and Au (lower panel) atoms intercalated in
MoS$_2$/Graphene heterostructure. 

\end{newpage}

\begin{newpage}
 
\begin{figure}
\includegraphics[width=8.5cm]{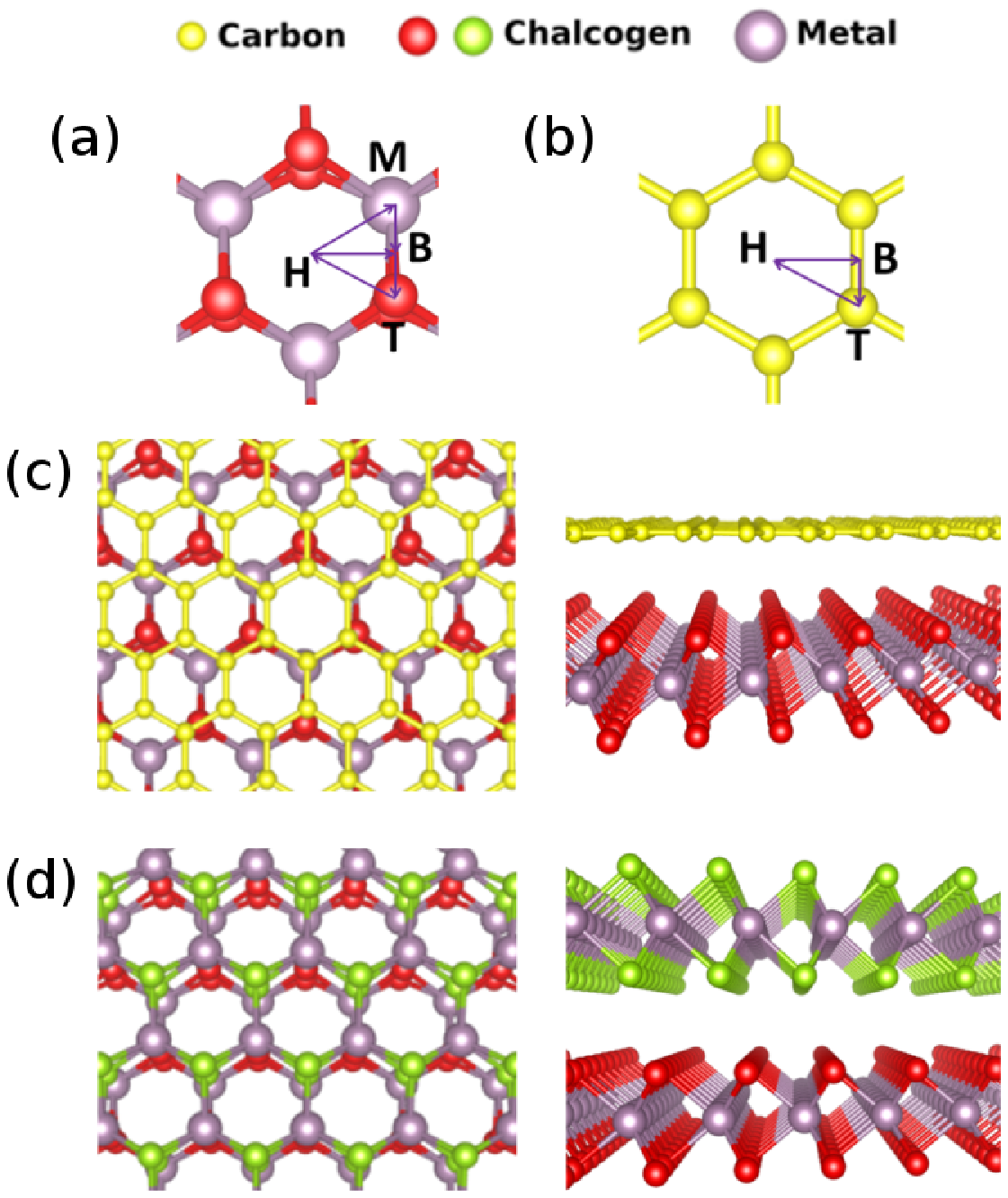}
\caption{\label{structures}
}
\end{figure}

\begin{figure}
\includegraphics[width=8.5cm]{fig2.eps}
\caption{\label{ag-hesap}
}
\end{figure}

\begin{figure}
\includegraphics[width=8.5cm]{fig3.eps}
\caption{\label{au-hesap}
}
\end{figure}

\begin{figure}
\includegraphics[width=8.5cm]{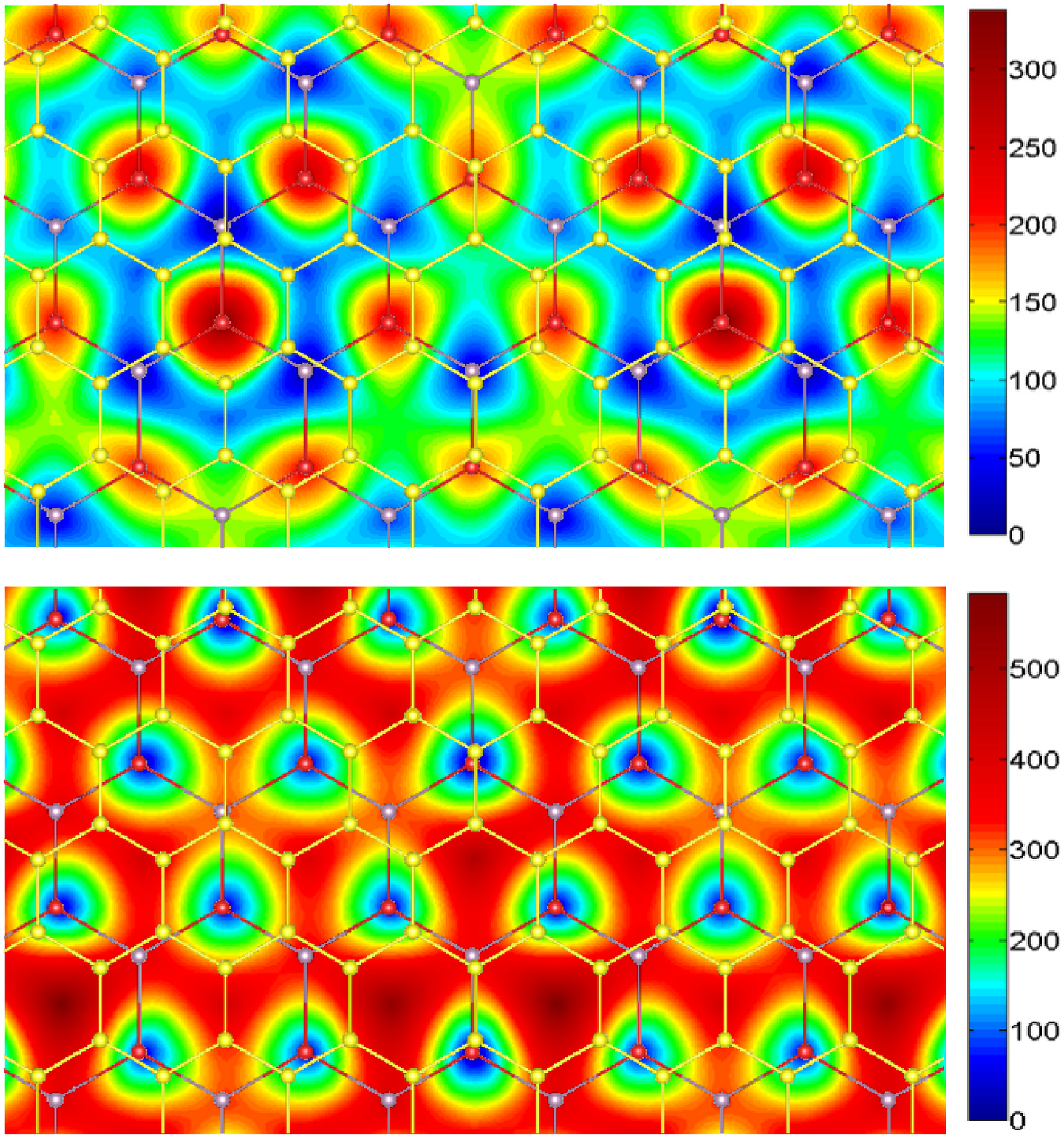}
\caption{\label{G/MoS2-Ag}
}
\end{figure}
 
\end{newpage}


\begin{thebibliography}{99}
%
 \bibitem{novoselov2004electric} K. S. Novoselov, A. K. Geim, S. V. Morozov, D. Jiang, Y. Zhang, S. V. 
 Dubonos, I. V. Grigorieva, and A. A. Firsov, Science \textbf{306}, 666 (2004).
%
 \bibitem{novoselov2005two} K. S. Novoselov, A. K. Geim, S. V. Morozov, D. Jiang, M. I. Katsnelson,
 I. V. Grigorieva, S. V. Dubonos, and A. A. Firsov, 
 Nature \textbf{438}, 197 (2005).
%
 \bibitem{bolotin2008temperature} K. I. Bolotin, K. J. Sikes, J. Hone, H. L. Stormer, and P. Kim, 
 Phys. Rev. Lett. \textbf{101}, 096802 (2008).
%
 \bibitem{novoselov2007room} K. S. Novoselov, Z. Jiang, Y. Zhang, S. V. Morozov, H. L. Stormer, U. Zeitler, 
 J. C. Maan, G. S. Boebinger, P. Kim, and A. K. Geim, 
 Science \textbf{315}, 1379 (2007).
%
 \bibitem{lee2008measurement} C. Lee, X. Wei, J. W. Kysar, and J. Hone, Science
\textbf{321}, 385 (2008).
%
 \bibitem{seol2010two} J. H. Seol, I. Jo, A. L. Moore, L. Lindsay, Z. H. Aitken, M. T. Pettes, X. Li, Z. Yao, 
 R. Huang, D. Broido, N. Mingo, R. S.  Ruoff, and L. Shi, 
 Science \textbf{328}, 213 (2010).
%
 \bibitem{sofo2007graphane} J. O. Sofo, A. S. Chaudhari, and G. D. Barber, Phys.
Rev. B 
 \textbf{75}, 153401 (2007).
%
 \bibitem{flores2009graphene} M. Z. S. Flores, P. A. S. Autreto, S. B. Legoas,
and D. S. Galvao, Nanotechnol. \textbf{20}, 465704 (2009).
%
 \bibitem{nair2010fluorographene} R. R. Nair, W. Ren, R. Jalil, I. Riaz, V. G. Kravets, L. Britnell, P. Blake, F. Schedin, A. S.  Mayorov, 
 S. Yuan, M. I. Katsnelson, H. M. Cheng, W. Strupinski, L. G. Bulusheva, A. V. Okotrub, I. V. Grigorieva, A. N. Grigorenko, K. S. Novoselov, 
 and A. K. Geim, 
 Small \textbf{6}, 2877 (2010).
%
 \bibitem{hasan-cf} H. Sahin, M. Topsakal, and S. Ciraci, 
 Phys. Rev. B \textbf{83}, 115432 (2011).
%
 \bibitem{ortwin-cf} H. Peelaers, A. D. Hernandez-Nieves, O. Leenaerts, B. Partoens, and F. M. Peeters, 
 Appl. Phys. Lett. \textbf{98}, 051914 (2011).
%
 \bibitem{hasan-ccl} H. Sahin, and S. Ciraci, J. Phys. Chem. C \textbf{116},
24075 (2012).
%
 \bibitem{kara2012review} A. Kara, H. Enriquez, A. P. Seitsonen, L. C. L. Y. Voon, S. Vizzini, B. Aufray, and H. Oughaddou, 
Surf. Sci. Rep. \textbf{67}, 1 (2012).
%
 \bibitem{PhysRevB.52.8881} H. Sahin, S. Cahangirov, M. Topsakal, E. Bekaroglu,
E. Akturk, R. T. Senger, and S. Ciraci Phys. Rev. B \textbf{80}, 155453 (2009).
%
 \bibitem{PhysRevB.8.3719} L. F. Mattheiss, Phys. Rev. B \textbf{8}, 3719
(1973).
%
 \bibitem{chhowalla2013chemistry} M. Chhowalla, H. S.  Shin, G. Eda, L. J. Li, K. P. Loh, and H. Zhang, 
 Nat. Chem. \textbf{5}, 263 (2013).
%
 \bibitem{yandong1} Y. Ma, Y. Dai, M. Guo, C. Niu, Y. Zhu, and B. Huang, 
 ACS Nano \textbf{6 (2)}, 1695 (2012).
 %
 \bibitem{hasan-nature}  S. Tongay, H. Sahin, C. Ko, A. Luce, W. Fan, K. Liu, J. Zhou, Y. S. Huang, C. H. Ho, J. Yan, 
 D. F. Ogletree, S. Aloni, J. Ji, S. Li, J. Li, F. M. Peeters, and J. Wu, 
 Nat. Commun. \textbf{5}, 3252 (2014).
%
 \bibitem{sipos2008mott} B. Sipos, A. F. Kusmartseva, A. Akrap, H. Berger, L. Forro, and E. Tutis, 
 Nat. Mater. \textbf{7}, 960 (2008).
%
 \bibitem{jishi2008electronic} R. A. Jishi, and H. M. Alyahyaei, 
 Phys. Rev. B \textbf{78}, 144516 (2008).
%
 \bibitem{splendiani2010emerging} A. Splendiani, L. Sun, Y. Zhang, T. Li, J. Kim, C. Y. Chim, G. Galli, and F. Wang,
Nano Lett. \textbf{10}, 1271 (2010).
%
 \bibitem{PhysRevB.88.075409} E. Cappelluti, R. Roldan, J. A. Silva-Guillen, P. Ordejon, and F. Guinea, 
 Phys. Rev. B \textbf{88}, 075409 (2013).
%
 \bibitem{hasan-wse2} H. Sahin, S. Tongay, S. Horzum, W. Fan, J. Zhou, J. Li, J. Wu, and F. M. Peeters, 
 Phys. Rev. B \textbf{87}, 165409 (2013).
%
 \bibitem{das2012high} S. Das, H. Y. Chen, A. V. Penumatcha, and J. Appenzeller, 
Nano Lett. \textbf{13}, 100 (2013).
%
 \bibitem{radisavljevic2011single} B. Radisavljevic, A. Radenovic, J. Brivio, V. Giacometti, and A. Kis, 
 Nat. Nanotechnol. \textbf{6}, 147 (2011).
%
 \bibitem{fang2012high} H. Fang, S. Chuang, T. C. Chang, K. Takei, T. Takahashi, and A. Javey, 
 Nano Lett. \textbf{12}, 3788 (2012).
 %
 \bibitem{geim2011random} A. K. Geim, 
 Intern. J. Mod. Phys. B \textbf{25}, 4055 (2011).
%
 \bibitem{geim2013van} A. K. Geim, and I. V. Grigorieva, 
 Nature \textbf{499}, 419 (2013).
%
 \bibitem{terrones2013novel} H. Terrones, F. L{\'o}pez-Ur{\'\i}as, and M. Terrones, 
Sci. Rep. \textbf{3}, 1549 (2013).
%
 \bibitem{haigh2012cross} S. J. Haigh, A. Gholinia, R. Jalil, S. Romani, L. Britnell, D. C. Elias, 
 K. S. Novoselov, L. A. Ponomarenko, A. K. Geim, and R. Gorbachev, 
Nat. Mat. \textbf{11}, 764 (2012).
%
 \bibitem{britnell2012field} L. Britnell, R. V. Gorbachev, R. Jalil, B. D. Belle, F. Schedin, 
 A. Mishchenko, T. Georgiou, M. I. Katsnelson, L. Eaves, S. V. Morozov, N. M. R. Peres, 
 J. Leist, A. K. Geim, K. S. Novoselov and L. A. Ponomarenko
 Science \textbf{335}, 947 (2012).
%
 \bibitem{dean2012graphene} C. Dean, A. F. Young, L. Wang, I. Meric, G. H. Lee, K. Watanabe, 
 T. Taniguchi, K. Shepard, P. Kim, and J. Hone, 
 Solid State Commun. \textbf{152}, 1275 (2012).
%
 \bibitem{georgiou2012vertical} T. Georgiou, R. Jalil, B. D. Belle, L. Britnell, R. V. Gorbachev, S. V. Morozov, Y. J. Kim, 
 A. Gholinia, S. J. Haigh, O. Makarovsky, L. Eaves, L. A. Ponomarenko, A. K. Geim, K. S. Novoselov and A. Mishchenko, 
 Nat. Nanotechnol. \textbf{8}, 100 (2012).
%
 \bibitem{yandong} Y. Ma, Y. Dai, W. Wei, C. Niu, L. Yu, and B. Huang,
 J. Phys. Chem. C \textbf{115 (41)}, 20237 (2011).
%
\bibitem{xinru} X. Li, Y. Dai, Y. Ma, S. Han, and B. Huang,
 Phys. Chem. Chem. Phys. \textbf{16}, 4230 (2014).
%
 \bibitem{joensen1986single} P. Joensen, R. F. Frindt, and S. R. Morrison, 
 Mat. Res. Bullet. \textbf{21}, 457 (1986).
%
 \bibitem{jaegermann1994photoelectron} W. Jaegermann, C. H. Pettenkofer, A. Schellenberger, C. A. Papageorgopoulos,
M. Kamaratos, D. Vlachos, and Y. Tomm, 
Chem. Phys. Lett. \textbf{221}, 441 (1994).
%
 \bibitem{tiong2001electrical} K. K. Tiong, Y. S. Huang, and C. H. Ho, 
 Journal of Alloys and Compounds \textbf{317}, 208 (2001).
%
 \bibitem{PhysRevB.78.134104} V. V. Ivanovskaya, A. Zobelli, A. Gloter, N. Brun, V. Serin, and C. Colliex, 
 Phys. Rev. B \textbf{78}, 134104 (2008).
%
 \bibitem{rao2012recent}  C. N. R. Rao, H. S. S. R. Matte, R. Voggu, and A. Govindaraj, 
Dalton Transactions \textbf{41}, 5089 (2012).
%
 \bibitem{shi2013selective} Y. Shi, J. K. Huang, L. Jin, Y. T. Hsu, S. F. Yu, L. J. Li, and H. Y. Yang, 
 Sci. Rep. \textbf{3}, 1839 (2013).
%
 \bibitem{kim2013enhanced} J. Kim, S. Byun, A. J. Smith, J. Yu, and J. Huang, 
 J. Phys. Chem. Lett. \textbf{4}, 1227 (2013).

%
 \bibitem{duygu} H. D. Ozaydin, H. Sahin, R. T. Senger, and F. M.
Peeters, Article first published online: 16 JUN 2014, DOI:
10.1002/andp.201400079


%
 \bibitem{matte2012synthesis} H. S. S. Matte, U. Maitra, P. Kumar, B. G. Rao, K. Pramoda, and C.N.R. Rao, 
 Zeitschrift f{\"u}r Anorganische und Allgemeine Chemie \textbf{638}, 2617
(2012).
%
 \bibitem{huang2010actively} F. Huang, and J. J. Baumberg, 
 Nano Lett. \textbf{10}, 1787 (2010).
%
 \bibitem{qian2012atomically} H. Qian, Y. Zhu, and R. Jin, 
 Proceedings of the National Academy of Sciences \textbf{109}, 696 (2012).
%
 \bibitem{he2012graphene} S. He, K. K. Liu, S. Su, J. Yan, X. Mao, D. Wang, Y. He, L. J. Li, 
 S. Song, and C. Fan, 
 Analytical Chem. \textbf{84}, 4622 (2012).
%
 \bibitem{lu2010platinum} Y. C. Lu, Z. Xu, H. A. Gasteiger, S. Chen, K. Hamad-Schifferli, and Y. Shao-Horn, 
 J. American Chem. Society \textbf{132}, 12170 (2010).
%
 \bibitem{Kretinin} A. V. Kretinin, Y. Cao, J. S. Tu, G. L. Yu, R. Jalil, K. S. Novoselov, S. J. Haigh, 
 A. Gholinia, A. Mishchenko, M. Lozada, T. Georgiou, C. R. Woods, F. Withers, P. Blake, G. Eda, A. Wirsig, C. Hucho, 
 K. Watanabe, T. Taniguchi, A. K . Geim, and R. V. Gorbachev,
 Nano Lett. \textbf{14 (6)}, 3270 (2014).
%
 \bibitem{Haigh} S. J. Haigh, A. Gholinia, R. Jalil, S. Romani, L .Britnell, D. C. Elias, K. S. Novoselov, 
 L. A. Ponomarenko, A. K. Geim, and R. Gorbachev, 
 Nat. Mater. \textbf{11}, 764 (2012).
%
 \bibitem{britnell2013strong} L. Britnell, R. M. Ribeiro, A. Eckmann, R. Jalil, B. D. Belle, A. Mishchenko, 
 Y. J. Kim, R. V. Gorbachev, T. Georgiou, S. V. Morozov, A. N. Grigorenko, A. K. Geim, C. Casiraghi, 
 A. H. C. Neto, and K. S. Novoselov, 
 Science \textbf{340}, 1311 (2013). 
%
 \bibitem{Sachs} B. Sachs, L. Britnell, T. O. Wehling, A. Eckmann, R. Jalil, B. D. Belle, A. I Lichtenstein, 
 M. I. Katsnelson, and K. S. Novoselov, 
 Appl. Phys. Lett. \textbf{103}, 251607 (2013).
%
 \bibitem{PhysRevB.54.11169} G. Kresse and J. Furthm\"uller, 
 Phys. Rev. B \textbf{54}, 11169 (1996).
%
 \bibitem{PhysRevLett.77.3865} J. P. Perdew, K. Burke, and M. Ernzerhof, 
 Phys. Rev. Lett. \textbf{77}, 3865 (1996).
%
 \bibitem{grimme2006semiempirical} S. Grimme, 
 J. Comp. Chem. \textbf{27}, 1787 (2006).
%
 \bibitem{amft} M. Amft, S. Lebegue, O. Eriksson, and N. V. Skorodumova,
 J. Phys. \textbf{23}, 395001 (2011).
%
 \bibitem{henkelman2006fast} G. Henkelman, A. Arnaldsson, and H. J{\'o}nsson, 
 Comp. Mat. Sci. \textbf{36}, 354 (2006).
%
 \bibitem{tang2009grid} W. Tang, E. Sanville, and G. Henkelman, 
 J. Phys.: Condensed Mat. \textbf{21}, 084204 (2009).
%
 \bibitem{sanville2007improved} E. Sanville, S. D. Kenny, R. Smith, and G. Henkelman, 
J. Comp. Chem. \textbf{28}, 899 (2007).
%

 \bibitem{Nakada} K. Nakada and A. Ishii, 
 ISBN 978 (2011).
%
 \bibitem{Yee} C. K. Yee, A. Ulman, J. D. Ruiz, A. Parikh, H. White, and M. Rafailovich, 
 Langmuir \textbf{19}, 9450 (1992). 
%

%
 \bibitem{Subrahmanyam} K. S. Subrahmanyam, A. K. Manna, and S. K. Pati 
 Chem. Phys. Lett. \textbf{497}, 70 (2010).
%
\bibitem{Bolla} B. G. Rao, H. S. S. R. Matte, and C. N. R. Rao, 
J. Cluster Sci. \textbf{23.3}, 929 (2012).
%
\bibitem{komsa2013electronic} H. P. Komsa and A. V. Krasheninnikov, 
 Phy. Rev. B \textbf{88}, 085318 (2013).
%
\bibitem{Gong} C. Gong, C. Huang, J. Miller, L. Cheng, Y. Hao, D. Cobden, 
J. Kim, R. S. Ruoff, R. M. Wallace, K. Cho, X. Xu and Y. J. Chabal, 
 ACS Nano \textbf{7}, 11350 (2013).


\end{thebibliography}
\end{document}